\documentclass[preprint,12pt]{elsarticle}

\usepackage{amssymb}
\usepackage{amsmath}

\usepackage{algorithm}
\usepackage{algpseudocode}
\usepackage{multirow}
\usepackage{array}
\usepackage{graphicx}
\usepackage{subcaption}
\usepackage{float}
\usepackage{xcolor} 
\usepackage[T1]{fontenc}
\usepackage[utf8]{inputenc}

\journal{International Journal of Electronics and Communications }

\begin{document}

\begin{frontmatter}



\title{An Efficient Algorithm for Modulus Operation and Its Hardware Implementation in Prime  Number Calculation}


\author{W.A.Susantha Wijesinghe} 
\address{{susantha@wyb.ac.lk},
            {}}

\address{{Department of Electronics},
	{Wayamba University of Sri Lanka},
	{Kuliyapitiya}, 
	{Sri Lanka}}

\begin{abstract}
This paper presents a novel algorithm for the modulus operation for FPGA implementation. The proposed algorithm use only addition, subtraction, logical, and bit shift operations, avoiding the complexities and hardware costs associated with multiplication and division. It demonstrates consistent performance across operand sizes ranging from 32-bit to 2048-bit, addressing scalability challenges in cryptographic applications. Implemented in Verilog HDL and tested on a Xilinx Zynq-7000 family FPGA, the algorithm shows a predictable linear scaling of cycle count with bit length difference (BLD), described by the equation $y=2x+2$, where $y$ represents the cycle count and $x$ represents the BLD. The application of this algorithm in prime number calculation up to 500,000 shows its practical utility and performance advantages. Comprehensive evaluations reveal efficient resource utilization, robust timing  performance, and effective power management, making it suitable for high-performance and resource-constrained platforms. The results indicate that the proposed algorithm significantly improves the efficiency of modular arithmetic operations, with potential implications for cryptographic protocols and secure computing.  
\end{abstract}

This is the accepted manuscript (post-peer-review, pre-typeset version) of the article published in International Journal of Electronics and Communications (AEÜ)\\ 

\vspace{0.5cm}
\textbf{Citation:} 
W.A. Susantha Wijesinghe,
An efficient algorithm for modulus operation and its hardware implementation in prime number calculation,
AEU - International Journal of Electronics and Communications,
Volume 191,
2025,
155657,
ISSN 1434-8411,\\
DOI: https://doi.org/10.1016/j.aeue.2024.155657

\vspace{0.5cm}
“© 2025. This manuscript version is made available under the CC BY-NC-ND 4.0 license [http://creativecommons.org/licenses/by-nc-nd/4.0/].”
\newpage
\begin{highlights}
\item A novel algorithm for modulus operations and its hardware implementation using only addition, subtraction, logical, and shift operations, avoiding complex multiplication and division.
\item Consistent performance across operand sizes from 32-bit to 2048-bit, with cycle count scaling linearly with bit length difference as y = 2x + 2.
\item Efficient FPGA implementation achieving 295.5 MHz maximum frequency for 2048-bit operations, with only $7,920$  LUT utilization and 0.174 W power consumption.
\item Application to prime number calculation demonstrating superior performance over software implementations, particularly for large integers.
\item High accuracy across operand sizes, with 100\% accuracy up to 2048-bit operations, suitable for cryptographic applications.

\end{highlights}


\end{frontmatter}



\section{Introduction}
\label{sec:introduction}

Residual Number System (RNS) continue to play a crucial role in hardware design for various applications, including cryptography, computer science, digital signal processing, error correction, and random number generation~\cite{oke2021residue}.  The efficiency of RNS implementations heavily relies on modular arithmetic operations.  While extensive research has been conducted on modular addition,  subtraction, multiplication, division, and exponentiation \cite{tynymbayev2019high},\cite{parihar2022low}, \cite{abd2021compact},\cite{ding2019low}, \cite{islam2020area}, and \cite{bos2020faster},  there is a notable gap in the literature regarding efficient hardware implementations of the fundamental modulus operation itself-that is, computing $A~mod~B$. 

The modulus operation, often overlooked in favor of its composite operations, is in fact the cornerstone of all modular arithmetic.  It is essential in cryptographic algorithms,  hash functions,  random number generation,  and error correction codes~\cite{sivakumar1995vlsi}\cite{muller2023area}.  Despite its importance, hardware implementations of the modulus operation have received comparatively little attention,  with most systems relying on  software implementations or treating it as a byproduct of division~\cite{will2016computing}.

This paper addresses this critical gap by presenting a novel algorithm specifically designed for modulus operation and its hardware implementation.  This focus on the fundamental operation of $A~mod~B$ distinguishes our work from the majority of studies in the field and offers several key advantages.  By optimizing the core modulus operation,  our work potentially improves the efficiency of all derived modular operations.  This cascading effect could lead to significant performance improvements in systems relying on modular arithmetic.  

The primary contributions of our research are as follows:
\begin{itemize}
	\item We present a novel algorithm for modulus calculation optimized for hardware implementation, addressing a fundamental operation often overlooked in hardware design literature.
	
	\item The algorithm is implemented in Verilog HDL and tested on a Xilinx Zynq-7000 family FPGA, demonstrating its practical applicability in real-world hardware environments.
	
	\item  Our algorithm can be easily implemented using a finite state machine (FSM) on FPGA hardware. By adjusting parameter values, the design can be scaled to any operand size, eliminating the need for complex designs.
	
	\item Our comprehensive evaluation reveals that the cycle count for the operation scales as $y = 2x + 2$, where $y$ represents the cycle count and $x$ represents the bit length difference  of the operands. 
	
	
	
	\item As a practical demonstration, we apply our novel algorithm for modulus operation to find prime numbers up to 500,000, showcasing its potential in cryptographic applications.
\end{itemize}

The rest of the paper is organized as follows: Section \ref{sec:relatedWork} summarizes the relevant work in the field.  Section \ref{sec:methodology} describes the algorithms and details of the implementations.  Results and Discussion are  presented in Section \ref{sec:results},  and Conclusion is provided in Section \ref{sec:concl}.

\section{Related Work}\label{sec:relatedWork}

This section provides an overview of relevant research in hardware implementations of modular operations,  highlighting the gap in literature regarding efficient hardware designs for the fundamental modulus operation. 

Modular arithmetic operations,  including addition,  subtraction,  multiplication and exponentiation,  have been the focus of numerous hardware implementation studies.  Styanarayana et al.  provided a comprehensive survey of hardware architectures for modular multiplication,  emphasizing its importance in public-key cryptosystems~\cite{Vollala2021}. Their work highlighted various reductions,  which are widely used but often involve complex hardware designs. 

Hossain et al.  presented hardware architectures for modular arithmetic operations over a prime field,  optimized for elliptic curve cryptography (ECC) \cite{hossain2019efficient}.  Their architectures focus on modular addition, subtraction,  and multiplication,  implemented separately to reduce circuit latency and area., achieving significant improvements in computational time and area utilization compared to related designs.  

Langhammer et al.  explored an efficient implementation of modular multiplication on FPGAs using Barrett's algorithm~\cite{langhammer2021efficient}.  Their method reduced resource count and latency of modular multiplication by employing aggressive truncation strategies for multipliers and introducing a new reduction method, demonstrating efficiency for 1024-bit modular multipliers.  However, the design introduced complexity in managing truncation errors and required significant DSP blocks. 

The growing importance of IoT and edge computing has sparked interest in efficient implementations of cryptographic operations on resource-constrained devices.  Ibrahim et al.  focused on the resource and energy-efficient hardware implementation of the Montgomery modular multiplication algorithm over $GF(2^m)$,  targeting compact IoT edge devices~\cite{ibrahim2023word}.  Their design achieved significant savings in area, delay, and energy consumption but involved complex scheduling and projection functions.

The modulus operation,  despite its fundamental nature,  has received relatively little attention in hardware design literature~\cite{butler2011fast}.  Sivakumar et al. explored VLSI architectures for computing the integer modulo operation $X~mod~m$ for specific values of $m$ \cite{sivakumar1995vlsi}. Their designs were optimized for specific modulus values, limiting general applicability.

Butler et al. presented a high-speed hardware implementation of the modulus operation optimized for FPGA deployment~\cite{butler2011fast}.  They introduced two versions of the algorithm to calculate $x~mod~z$: one with a fixed modulus and another where the modulus can vary. Their design demonstrated efficient pipelining and resource utilization but faced scalability issues for very large operand sizes. 

Will et al.   presented an efficient algorithm for modular reduction using a variable-sized lookup table, supporting large operands and relying on simple processor instructions, making it hardware friendly \cite{will2016computing}. However, the authors did not provide a hardware implementation, and the complexity of managing the lookup tables could be a limitation. 

Alia et al.  introduced a method for efficiently computing $x~mod~m$ using an approximation and correction approach that avoids direct division \cite{alia1990vlsi}, .  Their VLSI structure leveraged fast binary multipliers to handle 32-bit numbers. However, the complexity and resource requirements increased significantly for larger operands, limiting scalability. 

While the aforementioned studies have made significant contributions to the field of hardware-based modular arithmetic,  there remain a notable gap in the literature regarding efficient,  scalable hardware implementations of the fundamental modulus operation ($A~mod~B$).  The majority of existing research focuses on composite modular operations or overall cryptographic algorithms,  often overlooking the potential performance gains that could be achieved by optimizing this core operation.

\section{Methodology}\label{sec:methodology}

\subsection{Hardware Architecture for Modulus Operation}
The proposed algorithm computes modulus operation with addition,  subtraction,  logical,  and bit shift operations without using general multiplication or division operations which are expensive in hardware implementation.  The pseudocode of the proposed algorithm is given in Algorithm~\ref{alg:algorithm_mod}, which calculates $result = A~mod~B$, where $A$ is the dividend and $B$ is the divisor. 


\begin{algorithm}[!h]
\caption{\label{alg:algorithm_mod}The Novel Algorithm for Modulus Operation}	
\small
\begin{algorithmic}[1]
\Require N-bit integers $A$ and $B$
\Ensure N-bit integer $result$ where $result = \mod(A, B)$
\State Initialize $state \leftarrow IDLE$
\State Initialize $dividend \leftarrow 0$, $divisor \leftarrow 0$
\State Initialize $shift \leftarrow 0$
\State $done \leftarrow 0$
\While{true}
    \If{$state = IDLE$}
        \If{$start$}
            \State $dividend \leftarrow A$
            \State $divisor \leftarrow B$
            \State $shift \leftarrow 0$
            \State $state \leftarrow ALIGN$
        \EndIf
    \ElsIf{$state = ALIGN$}
        \If{$divisor \leq dividend$ and $\lnot divisor[N - 1]$ and $shift < N$}
            \State $divisor \leftarrow divisor \ll 1$
            \State $shift \leftarrow shift + 1$
        \Else
            \State $state \leftarrow SUBTRACT$
        \EndIf
    \ElsIf{$state = SUBTRACT$}
        \If{$dividend \geq divisor$}
            \State $dividend \leftarrow dividend - divisor$
        \EndIf
        \State $divisor \leftarrow divisor \gg 1$
        \State $shift \leftarrow shift - 1$
        \If{$dividend < divisor$ or $shift = 0$ or $shift \geq N$}
            \State $state \leftarrow FINISH$
        \EndIf
    \ElsIf{$state = FINISH$}
        \State $result \leftarrow dividend$
        \State $done \leftarrow 1$
        \State $state \leftarrow IDLE$
    \EndIf
\EndWhile
\end{algorithmic}
\end{algorithm}

The algorithm computes the $A~mod~B$ using state machine approach assuming that  $B>0$.  This assumption is validated at the system level during input checks before data is fed into the modulus hardware. It starts with initializing the \emph{state} to \emph{IDLE}  and setting variables \emph{dividend},  \emph{divisor},  \emph{shift},  and \emph{done} to zero.  The main loop continuously checks the \emph{state}.  If the \emph{state} is \emph{IDLE} and \emph{start} signal is received,  the algorithm assigns $A$ to the \emph{dividend} and $B$ to the \emph{divisor}, and the algorithm transitions to the \emph{ALIGN} state, where the \emph{divisor} is left-shifted until it aligns with the \emph{dividend},  incrementing the \emph{shift}  counter at each step. 

Once alignment is achieved,  the algorithm changes to \emph{SUBTRACT} state,  where it iteratively subtracts the \emph{divisor} from the \emph{dividend} if the \emph{dividend} is greater than or equal to the divisor.  The \emph{divisor} is then right-shifted,  and the \emph{shift} counter is decremented.  This process continues until the \emph{dividend} is smaller than $B$,  the \emph{shift} counter is zero,  or the \emph{shift} counter reaches $N$.  Finally, the algorithm changes to the \emph{FINISH} state,  where the \emph{result} is set to the current \emph{dividend},  and the state returns to \emph{IDLE},  indicating the operation is complete.  

\subsubsection{Numerical Example}
This section provides a detailed example of the the novel algorithm for modulus operation applied to two 6-bit integers,  ($A = 29$) and ($B = 5$), which computes the result of  $A \mod B $. This example illustrates how the algorithm operates by tracking changes in key variables: the $dividend$, $divisor$, and $shift$ counter, throughout the execution.

Let $ A = 29 $ and $ B = 5$ in a 6-bit system $ N = 6$. The initial values are:

\begin{align*}
    {Dividend} &= 29 \quad (\text{binary: } 11101_2) \\
    {Divisor } &= 5 \quad (\text{binary: } 00101_2) \\
    {Shift} &= 0
\end{align*}
We will now describe each step of the algorithm.\\
\textbf{Align the Divisor:}\\
To begin, the $divisor$ is left-shifted until it is greater than or equal to the $dividend$: 
\begin{align*}
    \text{Shift 1:} \quad & B \ll 1 = 10 \quad (\text{binary: } 00001010_2), \quad {Shift} = 1 \\
    \text{Shift 2:} \quad & 10 \ll 1 = 20 \quad (\text{binary: } 00010100_2), \quad {Shift} = 2 \\
    \text{Shift 3:} \quad & 20 \ll 1 = 40 \quad (\text{binary: } 00101000_2), \quad {Shift} = 3
\end{align*}
At this point, the $divisor$ ( 40 ) is greater than the $dividend$ ( 29 ), so the algorithm transitions to the subtraction phase.\\
\textbf{Subtract and Adjust:}\\
In this phase, the $divisor$ is right-shifted, and subtraction is performed when the $divisor$ is smaller than or equal to the $dividend$:
\begin{align*}
    \text{Shift right:} \quad & 40 \gg 1 = 20 \quad (\text{binary: } 00010100_2), \quad {Shift} = 2 \\
    \text{Subtract:} \quad & 29 - 20 = 9 \quad (\text{binary: } 01001_2), \quad {Shift} = 2 \\
    \text{Shift right:} \quad & 20 \gg 1 = 10 \quad (\text{binary: } 00001010_2), \quad {Shift} = 1 \\
    \text{Shift right:} \quad & 10 \gg 1 = 5 \quad (\text{binary: } 00000101_2), \quad {Shift} = 0 \\
    \text{Subtract:} \quad & 9 - 5 = 4 \quad (\text{binary: } 00100_2), \quad {Shift} = 0
\end{align*}
At the end of this phase, the $dividend$ is  4  and the $divisor$  ( 5 ) is now greater than the $dividend$. Therefore, the algorithm will finish with the remainder.\\
\textbf{Final Result:}
\[
result = dividend
\]
At this point, the $divisor$ has become greater than the $dividend,$ and the $shift$ counter has reached zero. Therefore, the algorithm terminates, and the final result of \( 29 \mod 5 \) is \( 4 \). 

\subsubsection{Hardware Implementation of the Modulus Operator}
The state machine approach ensures efficient and systematic computation of the modulus operation, as shown in Algorithm~\ref{alg:algorithm_mod}, making it well-suited for FPGA implementation. 

\begin{figure}[!h]
	\centering
	\includegraphics[width=0.6\linewidth]{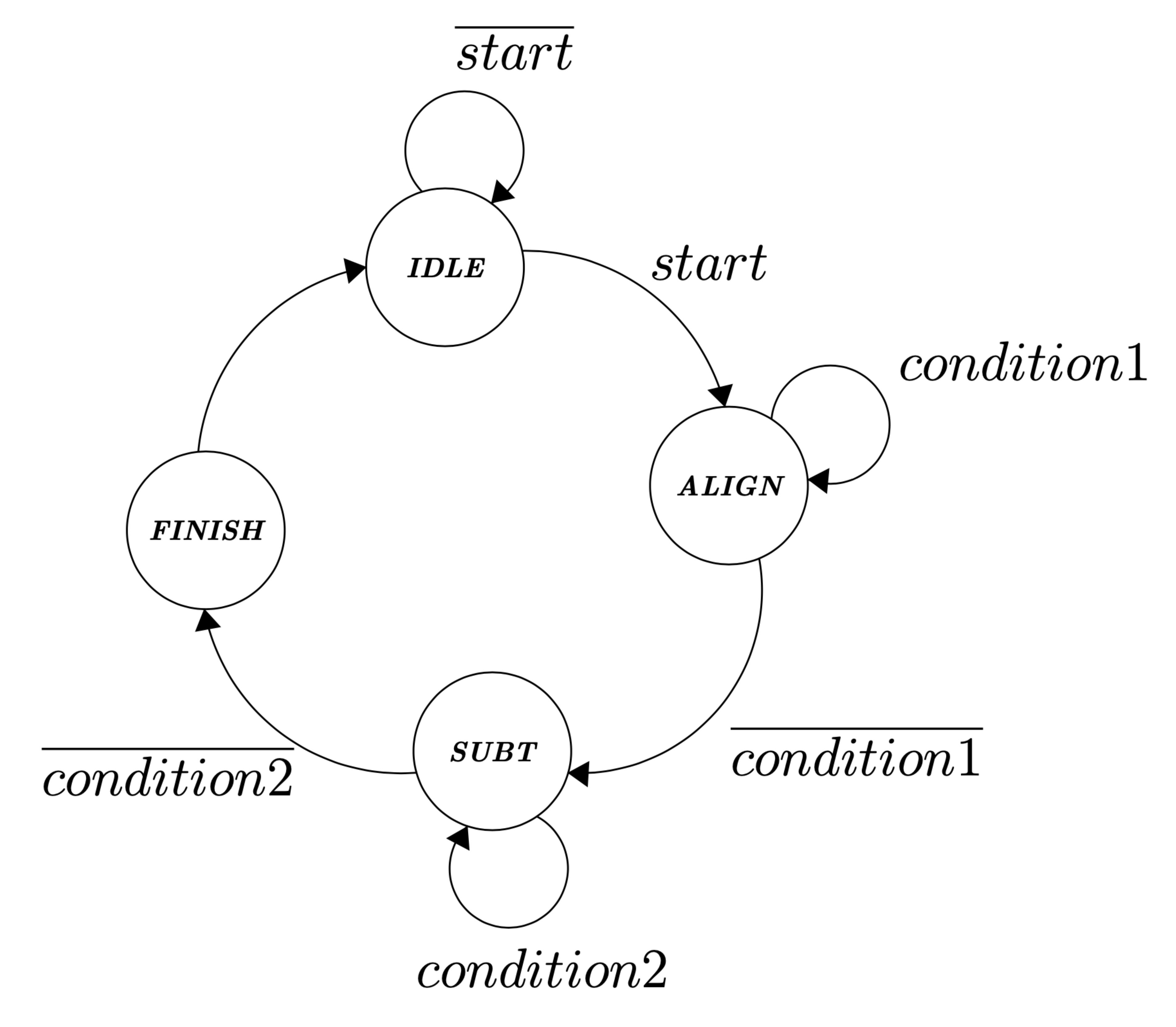} 
	\caption{\label{fig:fsm_modulus} Finite State Machine of the modulus algorithm.  Here
		$condition1 \equiv (divisor \leq dividend)~and~(!divisor[N-1]) ~and~(shift < N)$ and
		$condition2 \equiv (dividend < B)~or~(shift==0)~or~(shift \geq N) $}
\end{figure}

Figure~\ref{fig:fsm_modulus} presents the Finite State Machine (FSM) diagram with four states, operating as follows: Initially, the system remains in the \emph{IDLE} state until the \emph{start} signal is asserted. Once the \emph{start} signal is received, the system transitions to the \emph{ALIGN} state and stays there until \emph{condition1} is met; otherwise, it moves to the \emph{SUBT} state. In the \emph{SUBT} state, the system continues until \emph{condition2} is met, after which it transitions to the \emph{FINISH} state, signaling the completion of the calculation. The definitions of \emph{condition1} and \emph{condition2} are provided in the caption of Figure~\ref{fig:fsm_modulus}. The operations performed in each state are outlined in Table~\ref{tab:state_operations_fsm_modulus}.

\begin{table}[!h]
\caption{Operations assigned in each state of the FSM calculating the modulus operation}
\centering
\begin{tabular}{l | l}
\hline
\textbf{State} & \textbf{Operations} \\
\hline
\emph{IDLE} & 
\begin{tabular}[c]{@{}l@{}}
$dividend = A$ \\
$divisor = B$ \\
$shift = 0$
\end{tabular} \\
\hline
\emph{ALIGN} & 
\begin{tabular}[c]{@{}l@{}}
if $(divisor \leq dividend) ~and~!divisor[N-1]~ and~(shift < N)$: \\
\quad $divisor = divisor \ll 1$ \\
\quad $shift = shift + 1$
\end{tabular} \\
\hline
\emph{SUBT} & 
\begin{tabular}[c]{@{}l@{}}
if $dividend \geq divisor$: \\
\quad $dividend = dividend - divisor $\\
$divisor = divisor \gg 1$ \\
$shift = shift - 1$
\end{tabular} \\
\hline
\emph{FINISH} & 
\begin{tabular}[c]{@{}l@{}}
$result = dividend$ \\
$done = 1$
\end{tabular} \\
\hline
\end{tabular}
\label{tab:state_operations_fsm_modulus}
\end{table}

\begin{figure}[!h]
	\centering
	\includegraphics[width=0.8\linewidth]{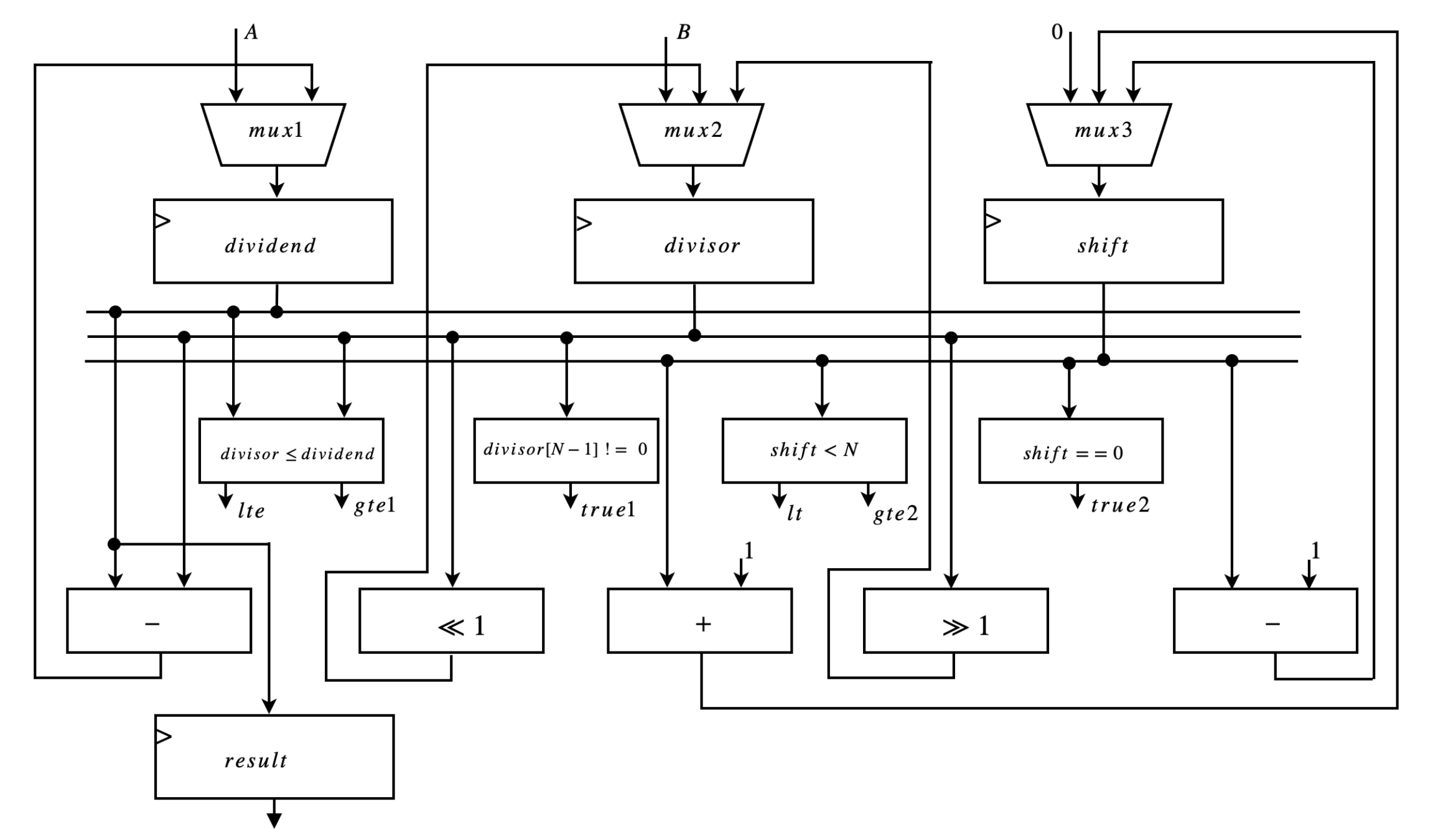} 
	\caption{\label{fig:RTLDiagramMod} Register Transfer Level (RTL) diagram illustrating the hardware architecture for implementing the proposed modulus operation algorithm on an FPGA.  }
\end{figure}

The Register Transfer Level (RTL) diagram shown in Figure~\ref{fig:RTLDiagramMod} illustrates the operations of the algorithm. Three registers are utilized: one for the dividend, one for the divisor, and one for shifting, which hold the current values. Each register receives its inputs through a separate multiplexer. Additionally, a distinct register is designated to capture the result of the computation. The comparator blocks perform the specified comparisons, while addition, subtraction, and shifting operations are executed on the current values. The outputs of these operations are connected to their respective multiplexers for updating the next values.

\begin{figure}[!h]
	\centering
	\includegraphics[width=0.8\linewidth]{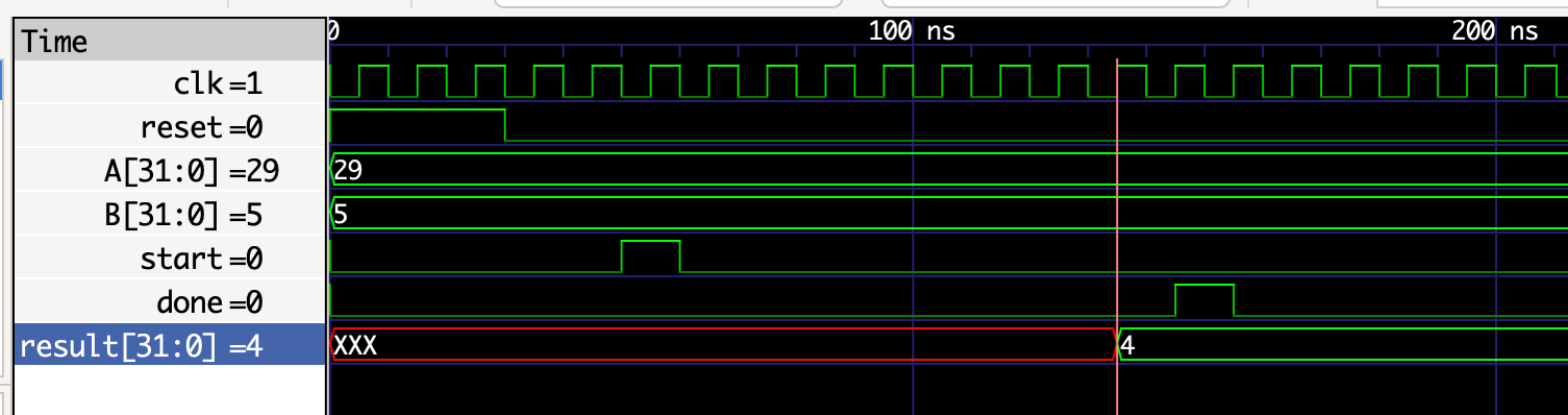} 
	\caption{\label{fig:simWaveformMod} The  waveform of the Verilog simulation of the proposed algorithm for modulus operation.  }
\end{figure}

\begin{figure}[!h]
	\centering
	\includegraphics[width=0.8\linewidth]{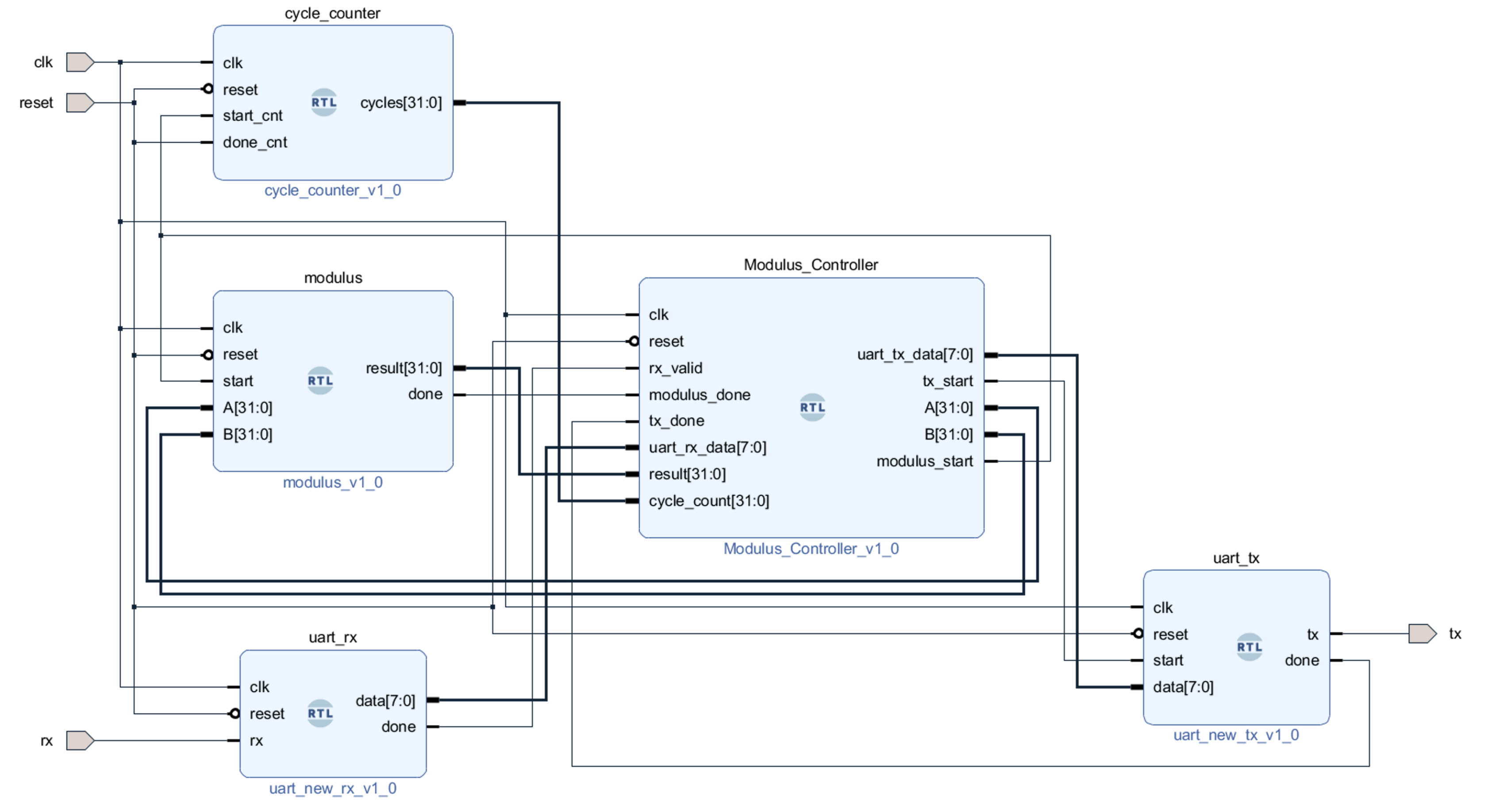} 
	\caption{\label{fig:blockDiagramMod} Block diagram illustrating the system for evaluating the modulus algorithm on the FPGA.  }
\end{figure}

The  algorithm for modulus calculation was described using the Verilog Hardware Description Language (HDL) to implement the FSM shown in Figure~\ref{fig:fsm_modulus}. The Xilinx Vivado design tool was employed for simulation and synthesis. Initially, the algorithm was tested with various input values to verify its accuracy in simulation. Figure~\ref{fig:simWaveformMod} shows the input/output waveforms from the simulation of finding \texttt{(29 mod 5)}, which is equal to 4. 

\sloppy
Following simulations, the Verilog implementation was expanded to include additional modules such as the UART transmitter(\texttt{uart\_tx}), UART receiver(\texttt{uart\_rx}), cycle counter(\texttt{cycle\_counter}), and controller module (\texttt{\linewidth=0.8\linewidth{Modulus\_Controller}}), as shown in the block diagram in Figure~\ref{fig:blockDiagramMod}.

The operation of the system to calculate $A \mod B$ is as follows: A Python program running on the computer sends the operands—dividend ($A$) and divisor ($B$) as 8-bit data packets through the UART transmitter to the FPGA board. The \texttt{uart\_rx} module receives the bitstream through the $rx$ input and outputs 8-bit data to the \texttt{Modulus\_Controller} module. The \texttt{Modulus\_Controller} module combines the 8-bit data received through \texttt{uart\_rx} to form the appropriate bit lengths for operands $A$ and $B$. Afterward, the \texttt{Modulus\_Controller} sends $A$ and $B$ to the \texttt{modulus} module and asserts the \texttt{modulus\_start} signal. At the same time, the \texttt{cycle\_counter} module begins counting clock cycles with the assertion of the \texttt{modulus\_start} signal. When the modulus calculation is complete, the \texttt{modulus} module generates the \texttt{modulus\_done} signal, and simultaneously, the \texttt{cycle\_counter} stops counting. The \texttt{Modulus\_Controller} retrieves the \texttt{result} from the \texttt{modulus} module and the cycle count from the \texttt{cycle\_counter}, then transmits this information through the \texttt{uart\_tx} module to the computer. The Python program running on the computer decodes the relevant data and stores it in a file.  The recorded data included the dividend,  divisor,  result,  and the number of cycles per calculation.  This comprehensive dataset was subsequently analyzed to evaluate the performance and efficiency of the modulus algorithm.

To measure the performance of the algorithm in FPGA hardware,  various operand sizes were used.  The algorithm can be extended to any bit length of operands by simply changing the integer value $N$ in the algorithm in Algorithm~\ref{alg:algorithm_mod}. For this study,  operand sizes of 32-bit,  64-bit,  128-bit,  256-bit,  1024-bit,  and 2048-bit were considered. 

\subsection{Application to Prime Number Calculation}

The best way to measure the performance of the novel algorithm for modulus operation is to use it in a practical application.  In this study,  we chose prime number calculation,  as it involves repeated modulus operations and is computationally intensive.  The application calculates all prime numbers up to a given integer.  For example,  if the given integer is 10,  the application finds 2,  3,  5, and 7 as prime numbers.  As the given integer increases,  the number of computations required grows significantly.  Therefore, this application provides an effective means to measure the performance of the new  algorithm for modulus operations.  

The pseudocode in Algorithm~\ref{alg:prime_algo} was constructed to identify all prime numbers up to a given integer $A$,  using only logical and addition operations. Although this prime number finding algorithm may not be the most optimized solution and there may be better alternatives,  it is sufficient for measuring the performance of the new modulus finding algorithm.

The algorithm begins by initializing several variables: $n$ (set to 1) iterate through numbers,  \emph{pCount} to count number of primes found,  $sel_1$ and $sel_2$ as selection flags,  and an empty list \emph{primes} to store the identified prime numbers.  The algorithm enters \emph{while} loop that continues as long as $n$ is less than $A$.  When $n$ equals 2,  the algorithm identifies it as a prime number, appends it to the \emph{primes} list,  and increments \emph{pCount}.  For numbers greater than 2,  the algorithm initializes $i$ to 3 and enters a nested \emph{while} loop that continues as long as $i$ is less than or equal to $n$.

Within the nested loop,  if $i$ equals $n$,  the number $n$ is identified as a prime,  appended to the \emph{primes} list,  and \emph{pCount} is incremented before breaking out of the loop. If $i$ is not equal to $n$,  the algorithm checks if $sel_2$ is 0,  in which case it calculates $s$ as the modulus of $n$ by 2 and sets $sel_2$ to 1.  Otherwise,  it calculates $s$ as the modulus of $n$ by $i$. If $s$ is not 0, $i$ is incremented by 2,  and the loop continues.  After exiting the nested loop,  $n$ is incremented by 1,  and the outer loop continues until $n$ reaches $A$.  The function finally returns a list containing the number of primes found and the list of prime numbers. 

	
\begin{algorithm}[H]
\caption{\label{alg:prime_algo}Prime number calculation up to a given integer $A$ using modulus operation}
\small
\centering
\begin{algorithmic}[1]
\State Initialize $n \gets 1$, $pCount \gets 0$, $sel_1 \gets 0$,  $sel_2 \gets 0$, $primes[]$
\While {$n < A$}
    \If {$n = 2$}
        \State $primes.append(n)$ 
        \State $pCount \gets pCount + 1$
    \Else
        \State $i \gets 3$
        \While {$i \leq n$}
            \If {$i = n$}
                \State $primes.append(n)$ 
                \State $pCount \gets pCount + 1$
                \State \textbf{break}
            \Else
            		\If {$sel2 = 0$}
                		\State $s \gets n \mod 2$
                		\State $sel2 \gets 1$
            		\Else
               		 \State $s \gets n \mod i$
                		\If{$s=0$}
                  		\State \textbf{break}
                		\EndIf
            		\EndIf
            \EndIf
                \State $i \gets i + 2$
        \EndWhile
    \EndIf
    \State $n \gets n + 1$
\EndWhile
\State \Return $[pCount, primes]$ 
\end{algorithmic}
\end{algorithm}

\subsubsection{Hardware Implementation of Prime Number Calculation}

\begin{figure}[!h]
	\centering
	\includegraphics[width=0.8\linewidth]{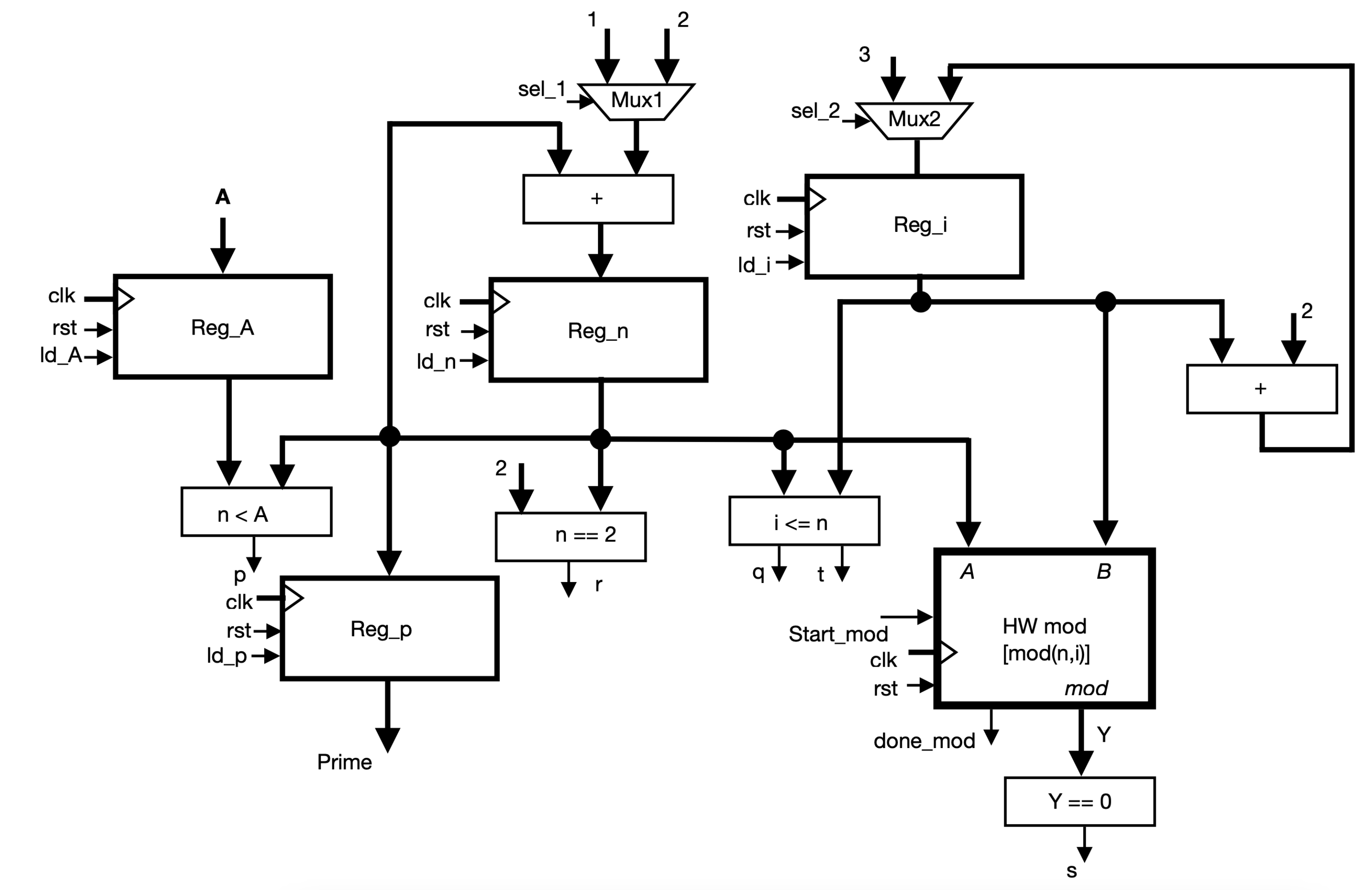} 
	\caption{\label{fig:primeAppDpath}Detailed data path diagram of the prime calculation system, including registers, the modulus module, and control signals.}
\end{figure}

Figure~\ref{fig:primeAppDpath} shows the datapath diagram of the prime calculation system. The system employs four registers to store the  input integer ($A$), two indexes ($n$ and $i$), and the prime number($P$). The signal $p$ is asserted when  $n < A$, and the signal $r$ is asserted when $n$ is equal to 2. Signals $q$ and $t$ are asserted when $i < n$ and $i$ is equal to $n$, respectively. The block \emph{mod} represents the hardware implementation of the modulus algorithm. 

The signal $s$ is asserted if the modulo operation  $n$ \emph{mod} $i$ equals zero. The signals \emph{start\_mod} and \emph{done\_mod}  indicate the initiation and completion of the modulo calculation process,  respectively.

\begin{figure}[!h]
	\centering
	\includegraphics[width=0.7\linewidth]{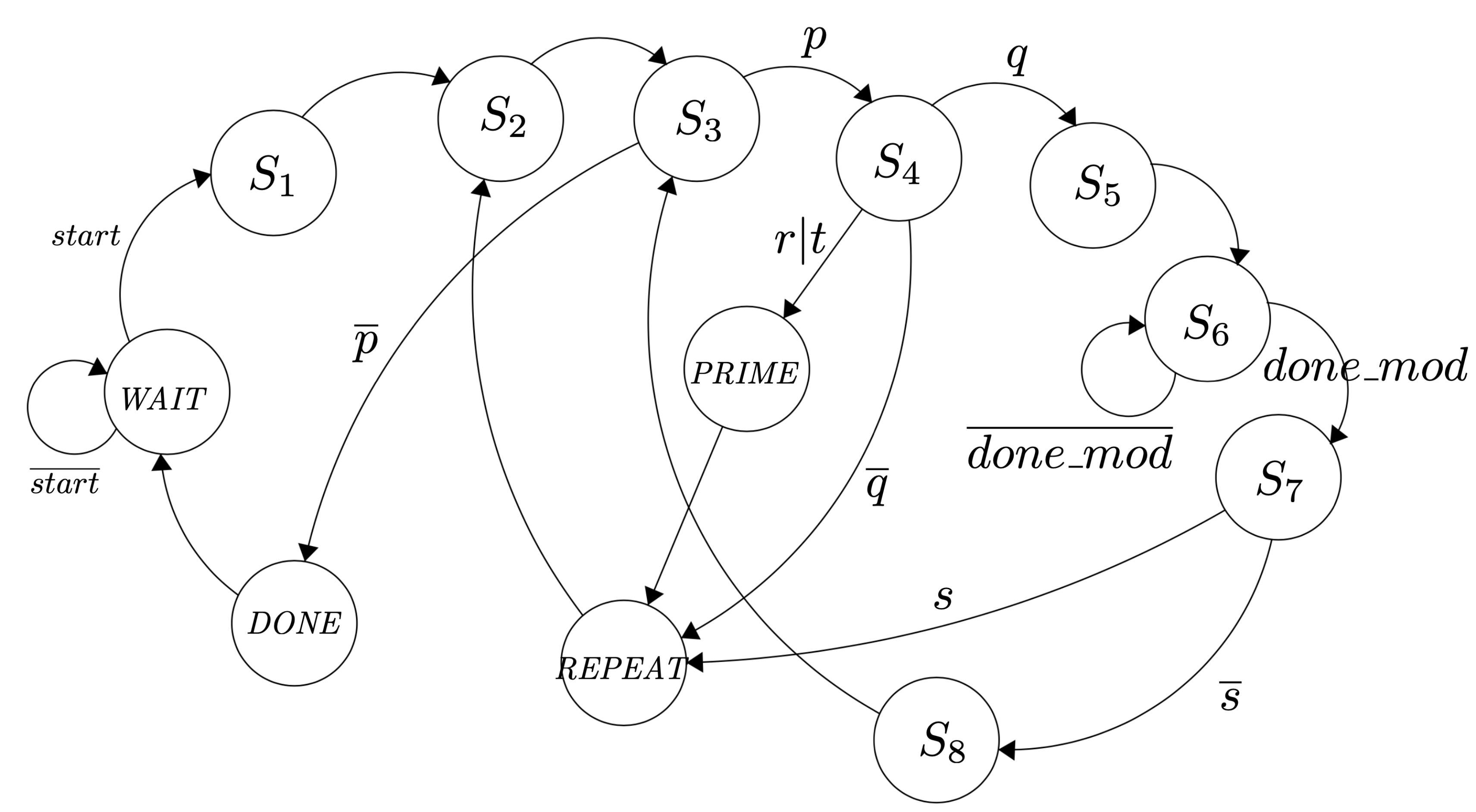} 
	\caption{\label{fig:FSM_diagramApp} Finite State Machine (FSM) diagram showing the control logic for the prime number calculation, outlining the various states and transitions involved in the process.}
\end{figure}

\begin{figure}[!h]
	\centering
	\includegraphics[width=0.6\linewidth]{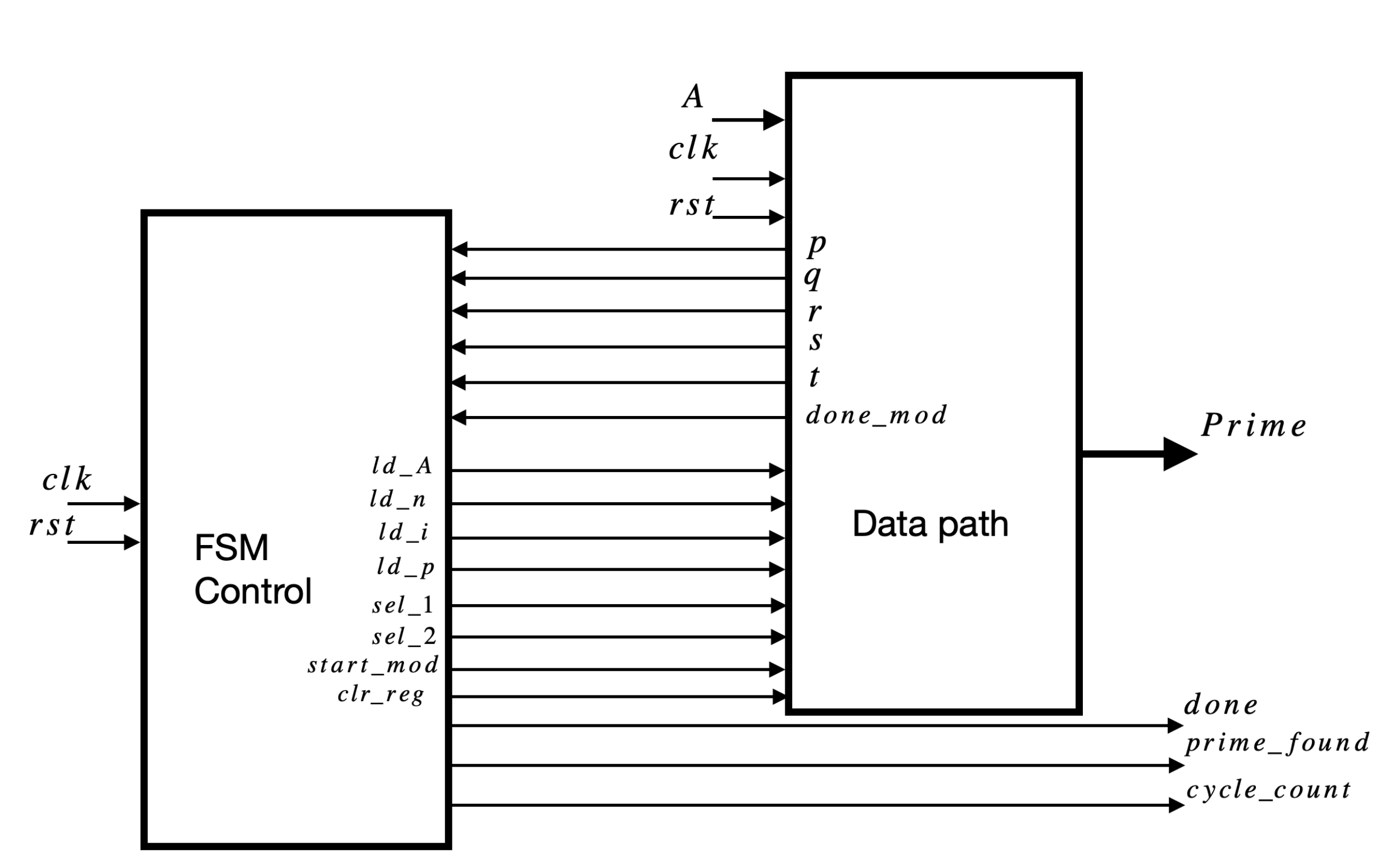} 
	\caption{\label{fig:fmodAppInterface}  Interface diagram between the data path and FSM controller, illustrating the interaction and signal flow necessary for prime number calculation.}
\end{figure}

Figure~\ref{fig:FSM_diagramApp} shows the finite state machine (FSM) diagram that generates the control signals to manage the data path.  The process begins with the \emph{start} signal, encompassing a total of 13 states. The control signals generated in each state are listed in Table~\ref{tab:StCtrlSiglsApp}. The datapath and the control FSM are implemented as separate Verilog modules and interfaced as  depicted in Figure~\ref{fig:fmodAppInterface}. Additionally, a counter module uses \emph{start},  \emph{prime\_found}, and \emph{done} signals to determine the number of primes found and the total clock cycles utilized for the entire process. 

Both hardware implementations, the modulus operation and the prime calculation, were configured using the Xilinx Vivado 2018.3 design suite,  targeting a Digilent Zybo Zynq-7000 SoC Trainer FPGA board containing a Xilinx XC7Z010-1CLG400C. Logic resource utilization,  power analysis,  and timing analysis were conducted using the Vivado design suite for both the hardware implementation of the novel algorithm for  modulus operation and the prime number calculation separately.

\begin{table}[!h]
	\centering
		\caption{\label{tab:StCtrlSiglsApp} Summary of control signals activated in each state of the FSM during the prime number calculation process, detailing the specific actions taken in each state.}

	\begin{tabular}{c | l }
		\hline
		\textbf{State} & \textbf{Control Signals} \\
		\hline
		\emph{WAIT}  & \emph{clr\_regs} = 1 \\
		\hline
		\multirow{2}{*}{$S1$} & $ld\_A$ = 1 \\
		&$ld\_n$ = 1 \\
		\hline
		\multirow{2}{*}{$S2$} &$ld\_n$ = 1\\
		&$sel\_2$ = 0\\ 
		\hline
		\multirow{2}{*}{$S3$}
		& if($sel\_2$ == 0)\\
		& \hspace{0.7cm} $ld\_i$ = 1\\
		\hline
		$S5$ & \emph{start\_mod} = 1 \\
		\hline 
		$S7$ & $sel\_2$ = 1 \\
		\hline 
		$S8$ & $ld\_i$ = 1 \\
		\hline 
		\multirow{2}{*}{\emph{PRIME}} & \emph{prime\_found} = 1\\
		& $ld\_P$ = 1 \\
		\hline 
		\multirow{4}{*}{\emph{REPEAT}} & if($r$ == 1)\\
		& \hspace{0.7cm} $sel\_1$ = 0 \\
		&else \\
		& \hspace{0.7cm} $sel\_1$ = 1\\
		\hline 
		\emph{DONE} & \emph{done} = 1\\
		\hline 
	\end{tabular}
\end{table}

The Verilog description of the datapath included several submodules,  such as registers and multiplexers,  in addition to the modulus module.  These modules were instantiated in the data path module along with other logical statements, following the datapath diagram shown in Figure~\ref{fig:primeAppDpath}.  The FSM was implemented using the two-always-block method in Verilog,  with one block for state logic and the other for next-state logic.  The top-level Verilog module integrates the FSM module and the datapath module.  The complete system was then simulated,  and the outputs were compared with software calculations to verify accuracy.

\subsection{Measurements}
In this study,  two main experiments were conducted to measure the performance of the proposed hardware algorithm for modulus operation and its application in finding prime numbers up to a given integer value.

To measure the accuracy and performance of the novel algorithm for modulus operation,  operand sizes of 32-bit,  64-bit,  128-bit,  256-bit, 1024-bit,  and 2048-bit were used.  Uniform random integers were generated through a Python program and encoded to send to the FPGA board via a UART interface for processing.  At the end of each operation,  results were sent from the FPGA board to the computer and recorded for further analysis.  For each operand size,  10,000 samples were recorded.

In the second experiment,  integer values of 10,  100,  1,000,  10,000,  100,000, 200,000,  300,000,  400,000,  and 500,000 were considered.  The prime numbers calculated up to each integer and the number of clock cycles used for each calculation were recorded.  These data were analyzed to measure the performance.

To compare the performance of the prime number calculations in FPGA hardware, software implementations of the prime number calculation algorithm (Algorithm~\ref{alg:prime_algo}) were considered.  In the software implementation of the algorithm,  the built-in modulus operator (\%) was used.  Both Python and C programming languages were used to run the same integer set on a Windows 11 PC with 8GB RAM and a processor with a 12th Gen Intel Core i5, 1300 MHz,  10 cores,  12 logical processors,  which can operate at a max turbo frequency of 4.9 GHz.  

\begin{figure}[h]
	\centering
	\includegraphics[width=0.6\linewidth]{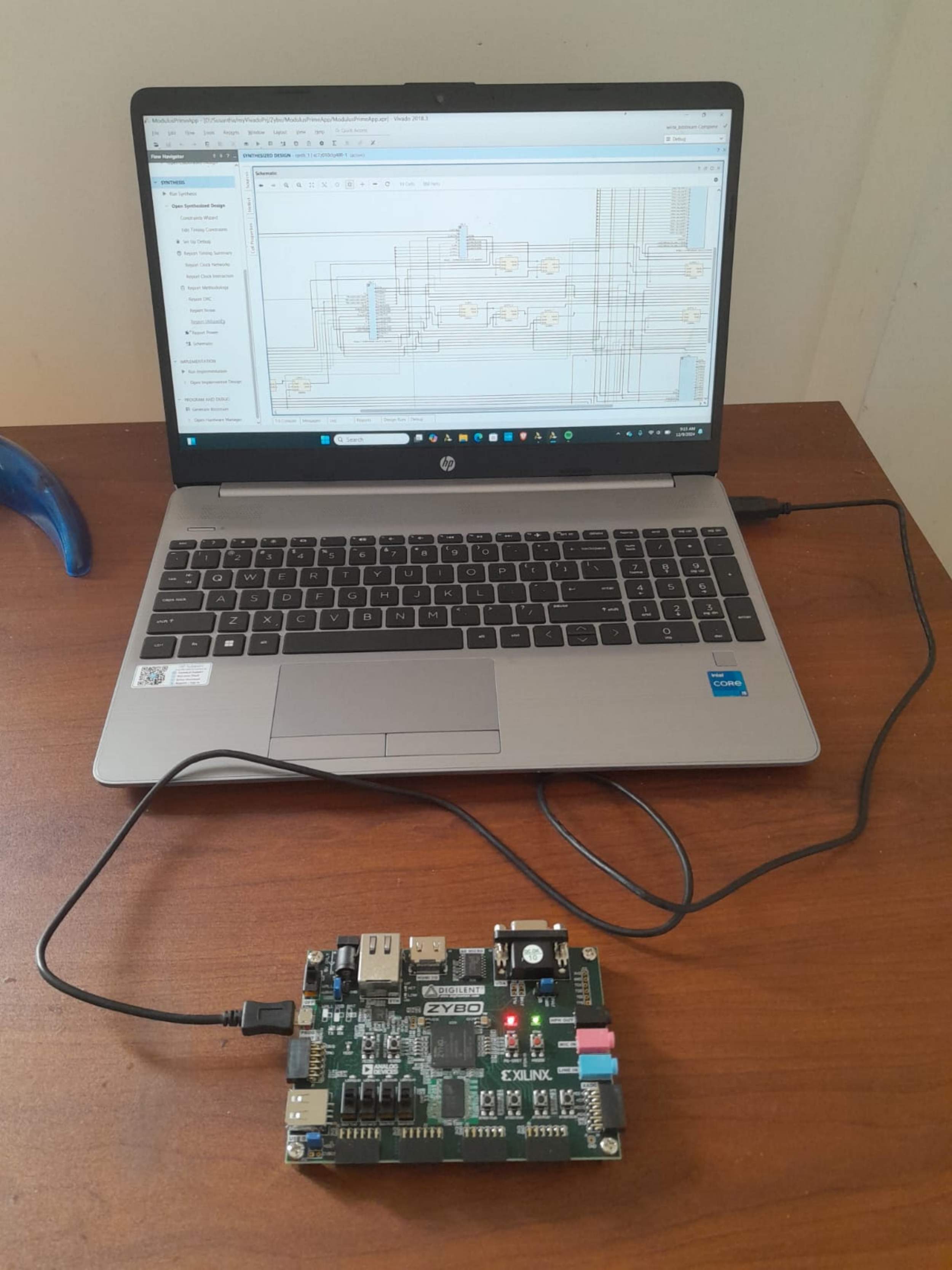} 
	\caption{\label{fig:image_hardware_setup} A picture of the hardware setup, where the Xilinx Zybo FPGA board is connected to the computer via the JTAG interface.  }
\end{figure}

Figure~\ref{fig:image_hardware_setup} shows an image of the hardware setup, where the Xilinx Zybo FPGA board is connected to the computer via the JTAG interface. This setup facilitates programming and debugging of the FPGA, while data transfer between the FPGA and the computer is managed through UART communication over the same JTAG interface, enabling real-time testing and performance evaluation.

\section{Results and Discussion}\label{sec:results}
\subsection{Hardware Implementation of the Novel Algorithm for Modulus Operation}
\begin{figure}[!h]
    \centering
    \begin{subfigure}[b]{0.48\textwidth}
        \centering
        \includegraphics[width=\textwidth]{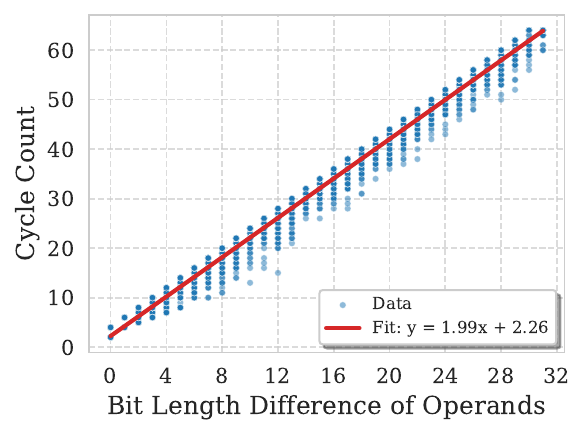}
        \caption{32-bit Operands}
        \label{fig:32bitOperands}
    \end{subfigure}
    \hfill
    \begin{subfigure}[b]{0.48\textwidth}
        \centering
        \includegraphics[width=\textwidth]{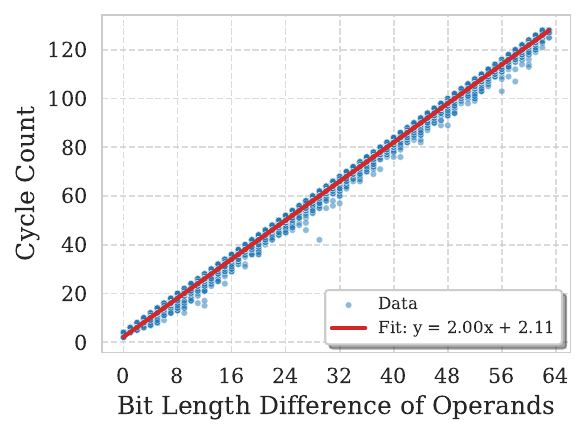}
        \caption{64-bit Operands}
        \label{fig:64bitOperands}
    \end{subfigure}
    
    \vspace{1em}
    
    \begin{subfigure}[b]{0.48\textwidth}
        \centering
        \includegraphics[width=\textwidth]{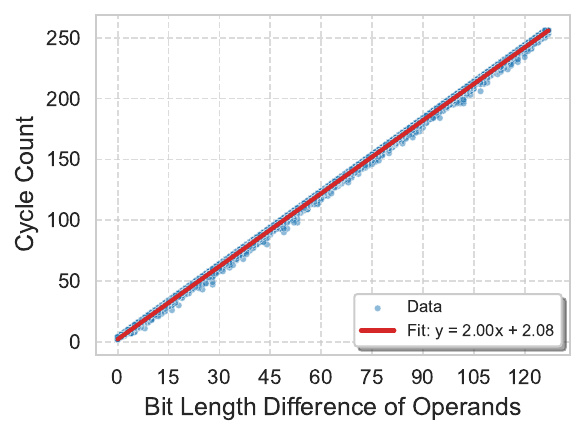}
        \caption{128-bit Operands}
        \label{fig:128bitOperands}
    \end{subfigure}
    \hfill
    \begin{subfigure}[b]{0.48\textwidth}
        \centering
        \includegraphics[width=\textwidth]{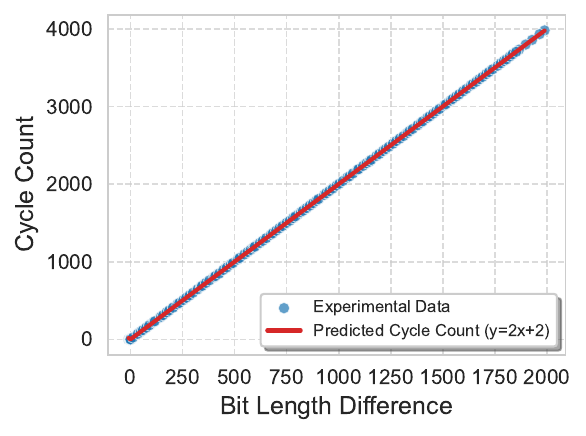}
        \caption{2048-bit Operands}
        \label{fig:2048bitOperands}
    \end{subfigure}
    \caption{Cycle Count vs Bit Lengths Difference of Operands.  Figure \ref{fig:32bitOperands}, \ref{fig:64bitOperands},  and \ref{fig:128bitOperands} show the experimental data and the linear fit for each operand size.  In Figure \ref{fig:2048bitOperands} shows the experimental data and the predicted cycle count by $y=2x + 2$.}
    \label{fig:modulus-results}
\end{figure}

Our hardware-implemented modulus algorithm achieved 100\% accuracy for all operand sizes when compared against software-based calculations.  The Figures~\ref{fig:32bitOperands}, ~\ref{fig:64bitOperands}, and~\ref{fig:128bitOperands} represent the performance of the hardware implemented algorithm for calculating the modulus operation with 32-bit, 64-bit, and 128-bit operands respectively.  

The equations of fits of three graphs approximately equivalent. The slopes of all three equations are nearly identical,  around 2.  This indicates that the number of cycles required for the modulus operation increases linearly with the bit length difference (BLD) at a consistent rate. The near-constant slope show that the algorithm's performance scales predictably regardless of the operand size,  making it robust and reliable. 

The $y$-intercepts are relatively low for all three cases. These values represent the base cycle count when the BLD is zero,  indicating minimal overhead.  The low $y$-intercept suggest that the algorithm starts with a small number of cycles and adds cycles linearly as the BLD increases,  highlighting the efficiency of the algorithm. Therefore,  we can say that cycle count and the BLD has the relationship shown in the equation~\ref{eq:cycleScale},  where $y$ represents cycle counts and $x$ represents BLD.
\begin{equation}\label{eq:cycleScale}
	y=2x +2
\end{equation}

The Figure~\ref{fig:2048bitOperands} illustrates the relationship between the the Cycle Count and the BLD for an experimental set of data where the operand size is 2048-bits.  The blue dots represent the experimental data points,  while the red line represents the predicted cycle count based on the equation~\ref{eq:cycleScale}. 

The red line closely follows the distribution of blue dots, indicating that the experimental data align well with the predicted cycle count. This highlights that the prediction equation, $y=2x+2,$ is highly accurate.

When the BLD is small, regardless of the operand sizes, the calculation completes in very few clock cycles. This is evident from the cluster of data points near the origin of the graph, where both BLD and cycle count are low. This indicates that the proposed novel algorithm performs very efficiently when the BLD is small. 

The hardware implementation of the modulus algorithm on the Zybo FPGA board, as analyzed using the Xilinx Vivado design environment, demonstrates efficient resource utilization, power consumption, and timing performance. 

The Table~\ref{table:resource_utilization} summarizes the resource utilization of the proposed  algorithm implemented on the Zybo FPGA board for different operand sizes (32-bit, 256-bit, 1024-bit, and 2048-bit). The key resources measured include Look-Up Tabes (LUT) and Flip-Flops (FF).  

As the operand size increases, the utilization of LUTs and FFs increase. This is expected as larger operand sizes require more complex logic. The utilization of BUFG remains constant at 6\% across all operand sizes. This indicates that the clock distribution requirements do not change with operand size.

\begin{table}[!h]
	\centering
		\caption{Resource utilization ( Look Up Tables-LUTs and Flip-Flops - FFs) summary for different operand sizes (32-bit, 256-bit, 1024-bit, and 2048-bit) on the Zybo FPGA.}
	\label{table:resource_utilization}
\resizebox{0.9\textwidth}{!}{
	\begin{tabular}{l c c c c c}
	\hline
	\multirow{2}{*}{\textbf{Resources} } & \multirow{2}{*}{\textbf{Available}} & \multicolumn{4}{c}		{\textbf{Utilization}} \\
	\cline{3-6}
	& & \textbf{32-bit} & \textbf{256-bit} & \textbf{1024-bit} & \textbf{2048-bit} \\
	\hline
	LUTs & 17,600 & 1,056(6\%) & 2,640(15\%) & 4,224(24\%) & 7,920(45\%) \\
	FFs & 35,200 & 1,760(5\%) & 3,168(9\%) & 3,872(11\%) & 5,280(15\%) \\
	\hline
	\end{tabular} 
}
\end{table}

The timing performance metics of the 2048-bit implementation on the Zybo FPGA board are summarized in Table~\ref{table:timing_performance}. The Worst Negative Slack (WNS) is 3.384 ns and it indicates the maximum amount by which the design meets the setup time requirements. Other metrics also indicate that the design meets all user-specified timing constraints. Based on the WNS, the maximum operating frequency is about 295.5 MHz. 

\begin{table}[!h]
	\centering
	\caption{Timing performance metrics for the 2048-bit implementation on the Zybo FPGA.}
	\label{table:timing_performance}
	\begin{tabular}{ l c }
		\hline
		\textbf{Metric} & \textbf{Value} \\ \hline 
		Worst Negative Slack (WNS) & 3.384 ns \\ 
		Maximum Frequency & 295.5 MHz\\ \hline 
	\end{tabular} 
\end{table}

The power performance metics are detailed in Table~\ref{tab:power_analysis}. The total on-chip power consumption is 0.174 W, divided into dynamic and static components. The dynamic power is 0.083 W, which constitutes the power consumed due to the switching activity of the circuit. Static power 0.091 W represents the power consumed due to leakage currents. The junction temperature is maintained at 27.0 C, with a thermal margin of 58.0 C, indicating effective thermal management. 

The 2048-bit implementation on the Zybo FPGA board demonstrates robust timing performance, with no violations in setup, hold, or pulse width requirements, and can operate at a maximum frequency of approximately 295.5 MHz. The power consumption is efficiently managed, with a balanced distribution between dynamic and static power components. These results validate the efficacy of the proposed \textcolor{blue}{algorithm } for high-performance applications on FPGA platforms. 

\begin{table}[!h]
	\centering
	\caption{Power consumption metrics for the 2048-bit implementation on the Zybo FPGA.}
	\begin{tabular}{l c}
		\hline
		\textbf{Parameter} & \textbf{Value} \\
		\hline
		Total On-Chip Power & 0.174 W \\
		Dynamic Power & 0.083 W (48\%) \\
		Device Static Power & 0.091 W (52\%) \\
		\hline
	\end{tabular} 
	\label{tab:power_analysis}
\end{table}

\subsection{Limitations of the Algorithm}
One limitation of our algorithm is that the cycle count required to complete modulus operations is not fixed, as it depends on the bit lengths of the dividend and the divisor. This variability can create challenges in determining the latency and throughput of designs. Another limitation is that the cycle count becomes significantly large when there is a considerable difference in the bit lengths of the operands. Addressing these limitations will be the focus of future improvements to the algorithm.

\subsection{Prime Number Calculation}

\begin{figure}[!h]
	\centering
	\includegraphics[width=0.6\linewidth]{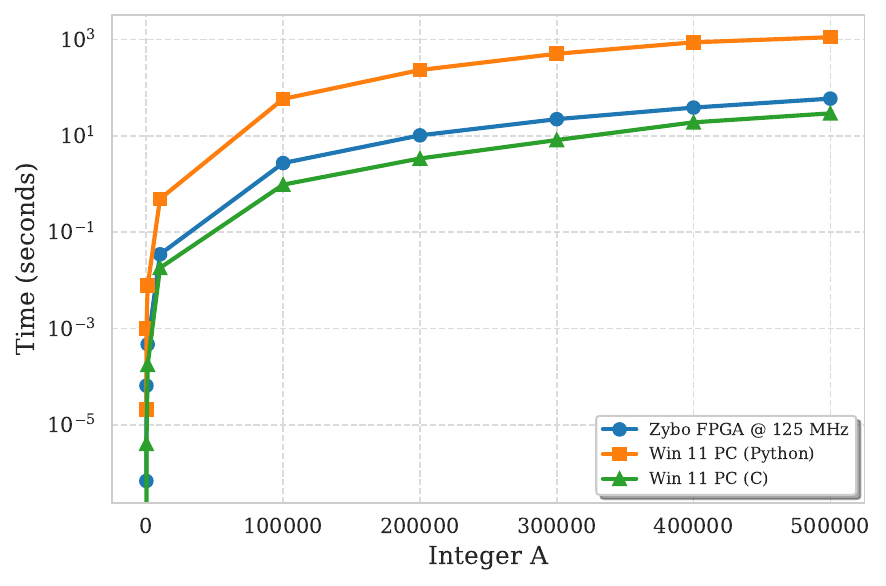}
	\caption{\label{fig:performanceCompare}Performance comparison of prime number calculation algorithms up to integer A. The graph shows computation times for hardware implementation on Zybo FPGA at 125 MHz, and software implementations in Python and C on a Windows 11 PC, illustrating the efficiency of the hardware approach, especially for larger integers. }
\end{figure}

Figure~\ref{fig:performanceCompare} illustrates the time taken to compute prime numbers up to a given integer A using three different approaches: a hardware implementation on the Zybo FPGA running at 125 MHz, and software implementations in Python and C on a high-performance Windows 11 PC. The y-axis is set to a logarithmic scale to highlight the differences in performance across several orders of magnitudes, while the x-axis represents the integer value for which primes are calculated.
Form the Figure~\ref{fig:performanceCompare} we can observe that the Python implementation shows significantly longer computation times compared to both the hardware and C implementations, while C implementation performs better than hardware implementation. Notably, the hardware implementation of the algorithm on Zybo FPGA is currently operating at 125 MHz, but it has potential to operate at a maximum frequency of 295.5 MHz. If the hardware were run at this higher frequency, the performance could potentially surpass that of the C implementation on the high-performance PC, particularly for lager integer values.

\begin{table}[!h]
	\centering
	\caption{Comparison of the time taken to find primes up to different integers using the Zybo FPGA at 125 MHz, Python, and C implementations on a high-performance PC. The table includes the number of primes found and the computation time for each method.}
	\small
	\begin{tabular}{r r r r r}
		\hline
		\textbf{Integer} & \textbf{Primes} & \textbf{Zybo FPGA} & \textbf{Win 11 PC} & \textbf{Win 11 PC} \\
		\textbf{A} & \textbf{found} & \textbf{@ 125 MHz} & \textbf{(Python)} & \textbf{(C)} \\
		& \textbf{up to A} & \textbf{Time (s)} & \textbf{Time (s)} & \textbf{Time (s)} \\
		\hline
		10 & 4 & 0.000000680 & 0.000020943 & $\approx{0.000000000} $ \\
		100 & 25 & 0.000064580 & 0.000998400 & 0.000040000 \\
		1,000 & 168 & 0.000468280 & 0.007677970 & 0.000176000 \\
		10,000 & 1,229 & 0.034511140 & 0.487490000 & 0.018000000 \\
		100,000 & 9,592 & 2.710378531 & 58.110090000 & 0.965000000 \\
		200,000 & 17,984 & 10.184829460 & 231.585353000 & 3.374000000 \\
		300,000 & 25,997 & 22.087260698 & 503.831983000 & 8.135000000 \\
		400,000 & 33,380 & 38.444923792 & 870.262950000 & 18.968000000 \\
		500,000 & 41,538 & 58.998911240 & 1112.491425000 & 29.179000000 \\
		\hline
	\end{tabular}
	\label{tab:prime-finder-performance}
\end{table}

\subsection{Comparison}

Butler et al. \cite{butler2011fast} presented results for computing x mod z in two scenarios: with z fixed at 3 and with z as a variable. For both cases, using 256-bit numbers for x, their system achieved an operating frequency of 143.8 MHz. The FPGA resource utilization was similar in both scenarios, consuming approximately 55,001 and 55,255 3-input LUTs for fixed and variable z, respectively.

In an earlier study, Alia et al. \cite{alia1990vlsi} proposed a method using $16 \times 16$ multipliers for calculating $x~mod~m$. Their implementation achieved a response time of 200 ns for 32-bit numbers, equivalent to an operating frequency of 5 MHz.

Our implementation of the modular exponentiation algorithm shows significant improvements over these previous works. For 2048-bit operand sizes, our design utilized only 7,920(45\%) of the available LUTs (17,600 total) on a Zybo FPGA board. Compared to Butler et al. \cite{butler2011fast}, our implementation uses substantially fewer FPGA resources while achieving a higher operating frequency of 295.5 MHz. This represents a notable advancement in both resource efficiency and performance for modular arithmetic operations on FPGAs.

In addressing the limitations found in the existing literature, our proposed algorithm offers several key improvements. One of the primary challenges identified in prior works is the complexity and resource intensity of performing molar arithmetic, particularly when relying on operations such as multiplication and division. Our approach, by focusing on addition, subtraction, logical operations and bit shifts, avoid these costly operations and thereby reduces both the hardware complexity and the associated resource utilization. Furthermore, the scalability of our algorithm is demonstrated by its linear cycle count scaling with operand size, an aspect often neglected in existing implementations where non-linear growth is observed.  This predictable scalability makes our design more suitable for cryptographic applications that require handling large operand sizes. Additionally, by optimizing the modulus operation itself, rather than focusing on composite operations such as modular multiplication or exponentiation, we provide a fundamental improvement that can be leveraged across modular arithmetic operations, further addressing the gaps in previous hardware designs.

\subsection{Discussion}

The linear relationship between the bit length difference (BLD) and the cycle count, as demonstrated in Figure \ref{fig:modulus-results}, highlights the efficiency of the proposed  algorithm for modulus operation. The algorithm scales predictably, maintaining efficiency even as operand sizes increase. This linear scalability is crucial for applications requiring high-speed computations with varying operand sizes. 

Figure~\ref{fig:2048bitOperands} provides additional validation of the algorithm's performance by comparing experimental data with the predicted cycle count. The close alignment of the experimental data points with the predicted cycle count  based on the equation $y=2x+2$ highlights the accuracy and robustness of the prediction model. The experimental results confirm that the  algorithm performs efficiently and predictably across different BLD values, maintaining low cycle counts even as the BLD increases. 

Analysis of FPGA resource utilization (Table \ref{table:resource_utilization}) demonstrates that our algorithm efficiently uses hardware resources even for large operand sizes, with the 2048-bit implementation using only 7,920(45\% of available) LUTs. Timing performance results (Table \ref{table:timing_performance}) confirm that this 2048-bit implementation meets all constraints without violations in setup, hold, or pulse width requirements, achieving a maximum frequency of 295.5 MHz. Furthermore, power consumption data (Table \ref{tab:power_analysis}) indicate a balanced distribution between dynamic and static components, with the total on-chip power maintained at an efficient 0.174 W.

The performance comparison of prime number calculations, illustrated in Figure~\ref{fig:performanceCompare}, shows that the hardware implementation on the Zybo FPGA outperforms the Python implementation and approaches the performance of the C implementation on a high-performance PC. Notably, the hardware implementation operates at 125 MHz, but with potential for higher performance at the maximum frequency of 295.5 MHz. This suggests that, with further optimization, the hardware implementation could surpass the software implementations, particularly for larger integer values.

The results validate the efficacy of the proposed  algorithm for modulus operations and its application in prime number calculations. The linear scalability, efficient resource utilization, robust timing performance, and effective power management make the algorithm well-suited for a wide range of hardware applications. Future work could explore further optimization to increase the operating frequency and reduce power consumption, as well as adaptations for other computationally intensive tasks.

\section{Conclusion}\label{sec:concl}
This study introduces a novel algorithm for the modulus operation, optimized for FPGA implementation. The algorithm's linear scalability and low overhead, demonstrated through extensive testing with operand sizes from 32-bit to 2048-bit, make it a robust solution for high-speed computations. The implementation on the Zybo FPGA platform confirms efficient resource utilization, robust timing performance, and effective power management, underscoring its suitability for both high-performance and resource-constrained platforms. Its application in prime number calculation further validates its practical use, showcasing substantial performance improvements over traditional software implementations. The findings suggest that this  algorithm can notably accelerate cryptographic protocols and other applications reliant on modular arithmetic, offering a promising direction for further research and development in hardware-based arithmetic operations. 




%
%
%


\end{document}